\documentstyle[pra,aps,twocolumn,psfig,epsf]{revtex}

\newcommand{\bra}[1]{\mbox{$\langle #1 |$}}
\newcommand{\ket}[1]{\mbox{$| #1 \rangle$}}
\newcommand{\bracket}[2]{\mbox{$\langle {{#1}} \mathrel{ | {\vphantom
        {{#1} {#2}}} \kern-\nulldelimiterspace} {{#2}} \rangle$}}
\newcommand{\proj}[1]{\ket{#1}\bra{#1}}
\newcommand{\Hil}{\mbox{$\cal H$}}
\newcommand{\al}{\mbox{$\alpha$}}

\newcommand{\ros}{\mbox{$\rho_s$}}

\begin{document}
\draft

\title{Separabilty and entanglement of composite quantum systems}
\author{Maciej Lewenstein and Anna Sanpera}
\address{CEA/DSM/DRECAM/SPAM, Centre d'Etudes de Saclay,  
\mbox{91191 Gif-Sur-Yvette,France.}}

\date{\today}

\maketitle

\begin{abstract}
We provide a constructive algorithm to find the best separable approximation to
an arbitrary density matrix of a composite quantum system of finite dimensions.
The method leads to a condition of separability and to a measure of
entanglement.

\end{abstract}

\narrowtext

Entanglement and nonlocality are some of the most emblematic concepts embodied
in quantum mechanics\cite{EPR}. The non-local character of an entangled system is
usually manifested in quantum correlations between subsystems that have interacted in the past but are not longer interacting. Furthermore, these concepts play a crucial role in quantum information theory\cite{bennett}.

From a formal point of view, a state of a composite quantum system  
is called  ``inseparable'' (or ``entangled'') 
if it cannot be represented as a tensor product of states of its subsystems. 
On the contrary, a density matrix $\rho$ 
describes a {\it separable}-state if it can be expressed 
as a finite\cite{pawel} sum of tensor products of its subsystems:
\begin{equation}
\rho_s=\sum_{i} p_i (\rho^A_{i}\otimes\rho^B_{i}...\otimes\rho^{N}_{i})
;\;\; 1\ge p_i\ge 0
\label{sepdef}
\end{equation}
where $\rho^A_{i},\rho^B_{i},..,\rho^N_i$ are density matrices 
describing subsystems $A,B,...,N$,  respectively and $\sum_ip_i=1$.  
Thus, separable states are
those than can be produced by N distant observers (Alice, Bob,..,Norberto) 
that prepare their states ($\rho^A_i, \rho^B_i,...,\rho^N_i$) independently,  following common instructions ($p_i$) from a source\cite{per}. 
Let us, for the moment, restrict ourselves to binary composite systems, i.e. $\Hil={\cal H}_A\otimes {\cal H}_B$.
Using the spectral decompositions
of  $\rho^A_{i}$ and $\rho^B_{i}$ it is easy to rewrite the Eq. (\ref{sepdef}) in the form
 \begin{equation}
\rho_s=\sum_{\alpha} \lambda_{\alpha} P_{\alpha}\;\; 1\ge \lambda_{\alpha}\ge 0; \,\, \sum_{\alpha}\lambda_{\alpha}=1,
\label{sepdef1}
\end{equation}
where $\alpha$ is
a multi-index running over all distinct eigenvectors of  the matrices $\rho^A_{i}\otimes\rho^B_{i}$, and
$P_{\alpha}$ are projectors onto product states, i.e.  $P_{\al}\equiv\proj{e,f}$ (where $\ket{e}\in \Hil_A$ and $\ket{f}\in\Hil_B$). Separable states, $\ros$, are thus 
mixtures of product states and as such their correlations are purely classical.

The distinction between entangled and separable states is well established
for pure stated: entangled pure states do always violate Bell inequalities\cite{bell}.
For mixed states, however, the statistical properties of
the mixture can hide the quantum correlations embodied in the system, 
making thus the distinction between separable
and entangled enormously difficult\cite{pop1,rhomho}. 
Besides the importance of the subject from a fundamental point of view, 
this distinction has also important consequences for quantum information theory. Consider, for instance,  Werner's family of entangled mixed states\cite{wer}, that does not violate any kind of Bell inequalities but, nevertheless,  can be used for quantum teleportation\cite{pop2}.

Recently, a first step in such distinction has been done by Peres\cite{per} and the Horodecki family\cite{pawel,ho3}. 
They have formulated two necessary conditions to characterize
separable density matrices.
The first condition\cite{per} 
states that if a matrix $\rho$ is separable, then its partial transposition (with respect to subsystem $A$, or $B$) must be a density matrix, i.e. must have
non-negative eigenvalues:
\begin{equation}
\rho=\ros\;\; \Rightarrow \rho^{T_B}=(\rho^{T_A})^{\star}\ge 0.
\label{partialpos}
\end{equation}
This can be easily grasped from the representation (\ref{sepdef1}) of separable matrices, since the partial transposition with respect to system B, amounts to replacing
$P_{\alpha}$ by $P^{T_B}_{\alpha}=|e,f^{\star}\rangle\langle e,f^{\star}|$, so that evidently
\begin{equation}
\label{rhotp}
\rho^{T_B}=\sum_{\alpha}\lambda_{\alpha}|e,f^{\star}\rangle\langle e,f^{\star}|\ge 0.
\end{equation}
This condition is sufficient to guarantee separability oly for composite systems of dimension $2\times 2$ or $2\times 3$. 

The second necessary condition\cite{pawel} 
states that if $\rho=\ros$, then there exist
a set of product vectors $V=\{|e_{i},f_{i}\rangle\}$ that  spans  ${\cal R}(\rho)$ and at the same time $V^{T2}=\{\ket{e_{i},f_{i}^{\star}}\}$ spans 
${\cal R}(\rho^{T2})$ where ${\cal R}(\rho)$ denotes the range of $\rho$,
i.e. the set of all $\ket{\psi}\in\Hil$ for which $\exists   \,\ket{\phi}\in\Hil$ such that $\ket{\psi}=\rho\ket{\phi}$.
From the representations (\ref{sepdef1}) and (\ref{rhotp}) we 
see that if a  set of product vectors $\{\ket{e_{i},f_{i}}\}$ spans
${\cal R}(\rho)$, it immediately follows that the set of  product vectors $\{|e_{i},f^{\star}_{i}\rangle\}$ also spans ${\cal R}(\rho^{T_B})$. 
In general, both conditions are not equivalent.
In particular, when the dimension of  ${\cal R}(\rho)$ is equal to the dimension of ${\cal R}(\rho^{T2})$, the second condition may not be
sufficient to ensure separability.

Finally, let us point out, that for a density matrix which is known to be separable, only if dim[$\Hil$] $\leq 6$ there exist an algorithm for decomposing  it according to Eq. (\ref{sepdef})\cite{rolf}.
 
In this Letter we address this last point and provide a constructive way
of finding such an algorithm regardless the (finite) 
dimension of the composite system. 
That immediately leads to a necessary condition for separability. 
Furthermore, we shall demonstrate that any
inseparable mixed state in ${\cal C}^2\otimes {\cal C}^2$ can be decomposed in a separable matrix and just a single pure entangled state, providing thus a novel characterization  of the ``entanglement'' of any inseparable state. 

The idea behind the algorithm relies on the fact that the set of separable states is compact. Therefore, for any density matrix $\rho$ there exist a ``maximal'' separable matrix $\rho_s^*$ which can be subtracted from $\rho$ maintaining the positivity of the difference, $\rho-\rho_s^*\geq 0$. 
Let us express the above idea in a more rigorous way:
\vskip 3mm
\noindent {\bf Theorem 1} For any density matrix $\rho$
(separable, or not) and for
any set $V$ of product vectors belonging to the range of $\rho$, i.e $|e,f\rangle\in {\cal R}(\rho)$
there exist a separable (in general not normalized) matrix 
\begin{equation}
\label{rho0}
\rho_s^*=\sum_{\alpha}\Lambda_{\alpha}P_{\alpha},
\end{equation}
with all $\Lambda_{\alpha}\ge 0$,
such that $\delta\rho=\rho-\rho_s^*\ge 0$, and that $\rho_s^*$ provides the best separable approximation (BSA) to $\rho$ in the sense that the trace ${\rm Tr}(\delta\rho)$ is minimal (or, equivalently, ${\rm Tr}\rho_s^*\le 1$ is maximal).
\vskip 2mm

The proof of the theorem is simple, and the whole art is, of course to construct $\rho_s^*$. Let us consider all separable matrices $\rho_s$ of the form (\ref{rho0})
that we can subtract from $\rho$ maintaining the non-negativity of the difference $\delta\rho$. Obviously, the trace of $\rho_s$ must be smaller
than one, since $0\le {\rm Tr}(\delta\rho)=1-{\rm Tr}\rho_s$. The set of such matrices is determined by the set of possible $\Lambda_{\alpha}\ge 0$ for
which  $\delta\rho\ge 0$, and $0\le{\rm Tr}\rho_s=\sum_{\alpha}\Lambda_{\alpha}\le 1$.
This set is closed (in any reasonable topology). The set of all possible traces of $\rho_s$ is bounded from above, so it must have an upper bound, say $1-\epsilon$; {\it ergo}  because of the compactness of the set of all $\rho_s$, there exist a matrix $\rho_s^*$ in this
set with the maximal trace, equal to $1-\epsilon$. 
That implies that although the matrix $\rho_s^*[V]$ depends on the choice of the set $V$, and by expanding $V$ we can construct better separable approximations to $\rho$ (i.e. for $V'\supset V$, ${\rm Tr}\rho_s^*[V']\ge {\rm Tr}\rho_s^*[V]$), it is generally sufficient to take $V\subset S$ large enough to obtain already the maximal possible trace  ${\rm Tr}\rho_s^*[V]={\rm Tr}\rho_s^*[S]$ (where $S$ is the set of all $\ket{e,f}\in{\cal R}(\rho)$). The latter statement indicates also that although  typically the BSA matrix $\rho_s^*[V]$ is not unique, its trace is. Nevertheless, for ${\cal C}^2\otimes{\cal C}^2$ composite systems we shall demonstrate that 
$\rho_s^*[V]$ is also unique.

As an obvious consequence of Theorem 1, we obtain a necessary and sufficient condition for separability:

\vskip 2mm
\noindent{\bf Condition 3} A density matrix $\rho$ is separable iff there exist 
a set of product vectors $V\subset {\cal R}(\rho)$, for which
the best separable approximation to $\rho$, $\rho_s^*[V]$ has the trace 1.
\vskip 2mm

The proof is again simple: The necessity of the cond3 follows directly from  (\ref{sepdef1}).
From the fact that 
$\delta\rho=\rho-\rho_s^*\ge 0$,
and  ${\rm Tr}\delta\rho=1-1=0$,  we obtain $\delta\rho\equiv 0$, or
equivalently $\rho=\rho_s^*$.

Before we discuss the procedure
of construction of the matrix $\rho_s^*$, let us to introduce two concepts which shall play a crucial 
role in what it follows.

\vskip 3mm
\noindent{\bf Definition 1} A non-negative parameter $\Lambda$ is called {\it maximal} with respect to a  (not necessarily normalised) density matrix
$\rho$, and the projection operator $P=|\psi\rangle\langle\psi|$ iff
$\rho-\Lambda P\ge 0$, and for every $\epsilon\ge 0$, the matrix $\rho-(\Lambda+\epsilon) P$ is not positive definite.

The maximal $\Lambda$ determines thus the maximal contribution of $P$ that can be subtracted from $\rho$ maintaining the non-negativity of the difference.
In the following we will apply the above definition to projections onto product vectors, i.e. $|\psi\rangle=|e,f\rangle$. The following lemma characterizes
a single maximal $\Lambda$ completely:
\vskip 3mm
\noindent{\bf Lemma 1}  $\Lambda$ is  maximal with respect to $\rho$ and $P=
|\psi\rangle\langle\psi|$ iff: (a) if $|\psi\rangle\not\in {\cal R}(\rho)$ then $\Lambda=0$, and (b) if $|\psi\rangle\in {\cal R}(\rho)$ then
\begin{equation}
0<\Lambda= \frac{1}{\langle \psi|\frac{1}{\rho}|\psi\rangle}.
\label{lamin}
\end{equation}
Note  that in the case (b) the expression on RHS of Eq. (\ref{lamin}) makes sense, since $|\psi\rangle\in {\cal R}(\rho)$, and therefore there exists $|\Psi\rangle\in {\cal R}(\rho)$ such that $|\psi\rangle=\rho|\Psi\rangle$. Let us observe, that
for any $|\phi\rangle$ the Schwartz inequality implies that
\begin{equation}
\langle\phi |P|\phi\rangle=
 |\langle \phi|\sqrt{\rho}\frac{1}{\sqrt{\rho}}|\psi\rangle|^2\le \langle\phi|\rho|\phi\rangle\langle\psi|\frac{1}{\rho}|\psi\rangle.
\end{equation}
That proves  that for every   $|\phi\rangle$, $\langle \phi|\rho -\langle\psi|1/\rho|\psi\rangle^{-1}P|\phi\rangle\ge 0$, i.e. $\rho-\Lambda P\ge 0$. Since on the other hand, $(\rho-\Lambda P)|\Psi\rangle=0$ for $|\Psi\rangle=\frac{1}{\rho}|\psi\rangle$,
thus for every $\epsilon>0$, $\langle\Psi|[\rho -(\Lambda+\epsilon)P]|\Psi\rangle=-\epsilon\Lambda^{-2}<0$. This proves that $\Lambda$ given by expression (\ref{lamin}) is indeed maximal.

\vskip 3mm
\noindent{\bf Definition 2} A pair of non-negative $(\Lambda_1,\Lambda_2)$
is called {\it maximal} with respect to $\rho$ and a pair of projection 
operators $P_1=|\psi_1\rangle\langle\psi_1|$, $P_2=|\psi_2\rangle\langle\psi_2|$
iff $\rho-\Lambda_1P_1-\Lambda_2P_2\ge 0$, $\Lambda_1$ is maximal with respect
to $\rho-\Lambda_2P_2$ and to the projector $P_1$, $\Lambda_2$ is maximal with respect to  $\rho-\Lambda_1P_1$ and to the projector $P_2$,  and the sum $\Lambda_1+\Lambda_2$ is maximal.

The maximal pair $(\Lambda_1,\Lambda_2)$ determines thus the maximal contribution of
$\Lambda_1P_1+\Lambda_2P_2$ that can be subtracted from $\rho$ maintaining the non-negativity of the difference, and that has a maximal trace,
${\rm Tr}(\Lambda_1P_1+\Lambda_2P_2)=\Lambda_1+\Lambda_2$. 

\vskip 3mm
\noindent{\bf Lemma 2} A pair $(\Lambda_1,\Lambda_2)$ is maximal with respect to $\rho$ and a pair of projectors $(P_1,P_2)$ iff: (a) if $|\psi_1\rangle, \;|\psi_2\rangle$ do not belong to ${\cal R}(\rho)$ then $\Lambda_1=\Lambda_2=0$;
(b) if $|\psi_1\rangle$ does not belong to
${\cal R}(\rho)$, while $|\psi_2\rangle\in {\cal R}(\rho)$
then $\Lambda_1=0$, $\Lambda_2=\langle \psi_2|1/\rho|\psi_2\rangle^{-1}$;
(c) if  $|\psi_1\rangle, \;|\psi_2\rangle \in {\cal R}(\rho)$ and 
$\langle \psi_1|1/\rho|\psi_2\rangle=0$ then $\Lambda_i=\langle \psi_i|1/\rho|\psi_i\rangle$, $i=1,2$; (d) finally,  if  $|\psi_1\rangle, \;|\psi_2\rangle \in {\cal R}(\rho)$ and 
$\langle \psi_1|1/\rho|\psi_2\rangle\ne 0$ then
\begin{mathletters}
\label{dupka}
\begin{eqnarray}
\Lambda_1&=&\left(\langle \psi_2|1/\rho|\psi_2\rangle-
|\langle \psi_1|1/\rho|\psi_2\rangle|\right)/D, \\
\Lambda_2&=&\left(\langle \psi_1|1/\rho|\psi_1\rangle-
|\langle \psi_1|1/\rho|\psi_2\rangle|\right)/D,
\end{eqnarray}
\end{mathletters}
where $D=\langle \psi_1|1/\rho|\psi_1\rangle\langle \psi_2|1/\rho|\psi_2\rangle
-|\langle \psi_1|1/\rho|\psi_2\rangle|^2$.
\vskip 2mm

The proof of (a) and (b) is the same as the proof of Lemma 1. In the case (c) observe that $(\rho-\Lambda_1P_1)^{-1}|\psi_2\rangle=\rho^{-1}|\psi_2\rangle$, 
 $(\rho-\Lambda_2P_2)^{-1}|\psi_1\rangle=\rho^{-1}|\psi_1\rangle$, so that
maximality of $\Lambda_i$ implies automatically that $\Lambda_i=\langle\psi_i|\rho^{-1}|\psi_i\rangle$, $i=1,2$.
Finally, in the case (d) we get  $(\rho-\Lambda_2P_2)^{-1}|\psi_1\rangle=\rho^{-1}|\psi_1\rangle + B\rho^{-1}|\psi_2\rangle$, with $B=\Lambda_2\langle \psi_2|1/\rho|\psi_1\rangle/D$. The maximality of $\Lambda_1$ assures
then automatically the maximality of $\Lambda_2$ provided
\begin{equation}
1-\Lambda_1\langle \psi_1|1/\rho|\psi_1\rangle-\Lambda_2\langle \psi_2|1/\rho|\psi_2\rangle+\Lambda_1\Lambda_2D=0.
\label{dupok}
\end{equation}
Maximizing the sum $\Lambda_1+\Lambda_2$ with the constraint (\ref{dupok}), 
we arrive after elementary algebra at Eqs. (\ref{dupka}). 

\noindent We can now formulate the basic theorem of this paper:
\vskip 3mm
\noindent{\bf Theorem 2} Given the set $V$ of product vectors $|e,f\rangle \in {\cal R}(\rho)$, the matrix 
$\rho_s^*=\sum_{\alpha}\Lambda_{\alpha}P_{\alpha}$ is the best
separable approximation (BSA) to $\rho$ iff
a) all $\Lambda_{\alpha}$
are maximal with respect to $\rho_{\alpha}=\rho-\sum_{\alpha'\ne\alpha}\Lambda_{\alpha'}P_{\alpha'}$,
and to the projector
$P_{\alpha}$; b) 
all pairs $(\Lambda_{\alpha},\Lambda_{\beta})$
are maximal with respect to $\rho_{\alpha\beta}=
\rho-\sum_{\alpha'\ne\alpha,\beta}\Lambda_{\alpha'}P_{\alpha'}$,  and to the
projection operators $(P_{\alpha},
P_{\beta})$.

\vskip 2mm
Let us prove now that maximizing all the pairs $(\Lambda_{\alpha},\Lambda_{\beta})$
with respect to $\rho_{\alpha\beta}=
\rho-\sum_{\alpha'\ne\alpha,\beta}\Lambda_{\alpha'}P_{\alpha'}$, $(P_{\alpha},
P_{\beta})$ is a necessary and sufficient condition to subtract the ``maximal'' separable matrix $\rho_s^*$ from $\rho$.
Obviously, if $\rho_s^*$ is the BSA then all $\Lambda_{\alpha}$, as well as 
all pairs $(\Lambda_{\alpha},\Lambda_{\beta})$ must be maximal, since otherwise
maximalizing $\Lambda_{\alpha}$, or the sum
$\Lambda_{\alpha}+\Lambda_{\beta}$ would increase the trace of $\rho_s^*$, maintaining non-negativity of $\rho-\rho_s^*$.

To prove the inverse, assume that the total number of $\alpha$'s is $K$,
and that $\rho_s^*$ has all pairs of $\Lambda$'s maximal. Consider matrices 
$\rho_s=\sum_{\alpha}\lambda_{\alpha}P_{\alpha}$ in the vicinity of $\rho_s^*$,
for which all individual $\lambda_{\alpha}$ are maximal, i.e.
$\rho_s$ belong to the boundary of the set $Z$ of all separable matrices 
such that $\rho-\rho_s\ge 0$; $\lambda_{\alpha}$'s lie thus on a $(K-1)$--dimensional manifold, defined through a constraint,
\begin{equation}
f(\lambda_1,\ldots,\lambda_K)=0.
\label{constr}
\end{equation}
Maximality of $(\Lambda_{\alpha},\Lambda_{\beta})$ implies that $(\lambda_{\alpha}+\lambda_{\beta})$ has a maximum at $\lambda_{\alpha,\beta}=
\Lambda_{\alpha,\beta}$ under the constraint (\ref{constr}), and for all
 $\gamma\ne \alpha,\beta$; $\lambda_{\gamma}=\Lambda_{\gamma}$ 
which implies ($\partial f/\partial \lambda_{\alpha}|_{\lambda=\Lambda})=
(\partial f/\partial \lambda_{\beta}|_{\lambda=\Lambda})$. Using this identity
for sufficient number of pairs we get that ($\partial f/\partial \lambda_{\alpha}|_{\lambda=\Lambda})={\rm const}$ for all $\alpha$. That is equivalent to the fact that the gradient of ${\rm Tr}(\rho_s)$ under the constraint (\ref{constr}) vanishes for $\rho_s=\rho_s^*$. The trace of $\rho_s$ has thus either a local maximum, or  a minimum, or a saddle point at $\lambda=\Lambda$. The two latter possibilities cannot occur, since the trace is maximal with respect to all pairs of $\lambda$'s, and since the set $Z$ is {\it convex} ( i.e. if $\rho_s,\rho_s'\in Z$ then 
$\epsilon\rho_s+(1-\epsilon)\rho_s'\in Z$ for every $0\le \epsilon\le 1$).
For the same reason of convexity, the local maximum at $\rho_s^*$ must be a 
global one, i.e. there cannot exist two matrices $\rho_s^*$, and $\tilde \rho_s^*$, which both provide local maxima of the trace, and have 
${\rm Tr}\rho_s^*\ne {\rm Tr}\tilde\rho_s^*$; {\it ergo} $\rho_s^*$ is the BSA,
and any other matrix $\tilde\rho_s^*$ which has all pairs of $\Lambda$'s maximal, must have the same trace as $\rho_s^*$.\\

In any case, we have shown that any density matrix $\rho$ of composite system $\Hil$ can be decomposed according to 
$\rho=\rho_s^*+\delta\rho$ where $\rho_s^*$ is a separable matrix (in general not normalized) with maximal trace. Let us analyze such
decomposition in more detail.
All the information concerning ``inseparability'' is included in the matrix $\delta\rho$. If it
does not vanish, i.e.  if $\rho$ is not separable,
its range ${\cal R}(\delta\rho)$ cannot contain 
any product vector.  We have observed that,
quite typically,  if $\delta\rho$ is
a sum of projections onto a set of linearly independent entangled states, then there exist product vectors that belong to ${\cal R}(\delta\rho)$, whose contributions can
be single out increasing ${\rm Tr \ros^*}$. The reason is
that, for instance, the set of all product vectors in the Hilbert space $H$ of dimension $N\times M$ spans a
$(N+M-1)$-dimensional manifold, which generically has a non-vanishing intersection with linear subspaces of $\Hil$ of dimension equal or larger than $(N-1)\times(M-1)$. The above statement implies that for $N=M=2$, $\delta\rho$ is a simple 
projector onto an entangled state. 

As an immediate consequence, we 
obtain that any density matrix $\rho$ in ${\cal  C}^2\otimes {\cal C}^2$ has a unique decomposition in the form:
\begin{equation}
\rho=\lambda{\ros}+(1-\lambda) P_e;\;\;\; \lambda\in[0,1]
\label{unique}
\end{equation}
where $\ros$ is a separable density matrix (normalized), $P_e$ denotes a single pure entangled projector ($P_e\equiv\ket{\Psi_e}\bra{\Psi_e}$), and $\lambda$ is maximal. Any other decomposition of the form $\rho=\tilde \lambda{\tilde \rho_s}+(1-\tilde \lambda)\tilde P_e$ with $\tilde\lambda\in[0,1]$ such that  $\tilde\rho_s\not=\ros$, necessarily implies that  $\tilde\lambda<\lambda$. If not, that is, if $\lambda=\tilde\lambda$ for $\tilde\rho_s\not=\ros$, it follows from Ref.\cite{rolf} that for $P_e\not=\tilde P_e$, 
we can always find projectors onto product states in the plane formed by $P_e$ and $\tilde P_e$ and therefore increase $\lambda$, which is impossible since
$\lambda$ is already maximal.

The decomposition given by expression (\ref{unique}) leads straightforwardly to an unambiguous measure of the entanglement for any mixed state $\rho$ (in ${\cal  C}^2\otimes {\cal C}^2$):
\begin{equation}
E(\rho)=(1-\lambda)\times E(\ket{\Psi_e})
\label{entan}
\end{equation} 
where  $E(\ket{\Psi_e})$ is the entanglement of its pure state expressed in
terms of the von Neumann entropy of the reduced density matrix of either of its
subsystems\cite{bennett2}:

\begin{equation}
E(\ket{\Psi_e})=-{\rm Tr}\rho_A\log_2\rho_A\equiv-{\rm Tr}\rho_B\log_2\rho_B
\label{pentan}
\end{equation}
where $\rho_{\{A,B\}}={\rm Tr}_{\{B,A\}}\rho$. This measure of entanglement is clearly independent of any purification or formation procedure\cite{bennett2,martin}.

Let us illustrate with an example the ideas stressed in the paper.
Consider a pair of spin-$\frac{1}{2}$ particles in an impure state consisting of a  fraction $x$ of the singlet and a mixture in equal proportions  of the singlet and the triplet\cite{wer}. This state is described, in the computational basis, by the density matrix:
\begin{equation}
\rho_{w}(x)=\pmatrix{\frac{1-x}{4} & 0 & 0 &  0\cr
                    0 & \frac{1+x}{4} & -\frac{x}{2}& 0\cr
		    0 & -\frac{x}{2} & \frac{1+x}{4}& 0\cr
		    0 & 0 & 0 &  \frac{1-x}{4}\cr}
;\;\; 0< x < 1
\label{wernma}
\end{equation}
For this case Eq.(\ref{partialpos}) is sufficient to ensure separability: $\rho_{w}$ is separable if $x\leq 1/3$ and
inseparable otherwise. Nevertheless, we use our procedure to check the separability and to obtain the decomposition of $\rho$ given by Eq. (\ref{unique}) for different values of $x$. 

For each given set $V$, we first construct the matrix
\begin{equation}
\rho_s^*[V]=\sum_{V}\Lambda_{\alpha} P_{\alpha}
\label{decom}
\end{equation}
with the $\Lambda '$ maximized pairwise, according to the definitions\cite{notka}. When the numerical convergence
has been achieved we obtain $\delta\rho=(\rho_{w}-\rho_s^*[V])$ 
and compute its trace. Typically, we observe that: (a) only very few projectors
$P_{\alpha}$ of each set $V$ contribute to the matrix $\rho_s^*[V]$, and (b) 
if the set $V$ is large enough (i.e $> 300$), the results become independent of the chosen set. 

The results are presented in Fig.1, for a set of 100, 200 and 500 $P_{\alpha}$-projectors randomly chosen. Each point represents the corresponding value of ${\rm Tr}(\delta\rho)$ for
a given $\rho_{w}(x)$. The vertical line indicates the condition of separability, derived
from Eq.(\ref{partialpos}). For $x\leq 1/3\;,$ ${\rm Tr}(\delta\rho)=0$ indicating that $\rho_w$ is separable. At $x\sim 1/3$, a clear ``phase-transition" occurs, and the value  ${\rm Tr}(\delta\rho)\not=0$, indicating thus the non-separable character of the state. Therefore, our numerical results reproduce accurately 
the conditions of separability derived from  Eq.(\ref{partialpos}).
\begin{figure}[h]
\centerline{
\psfig{angle=90,height=58mm,width=72mm,file=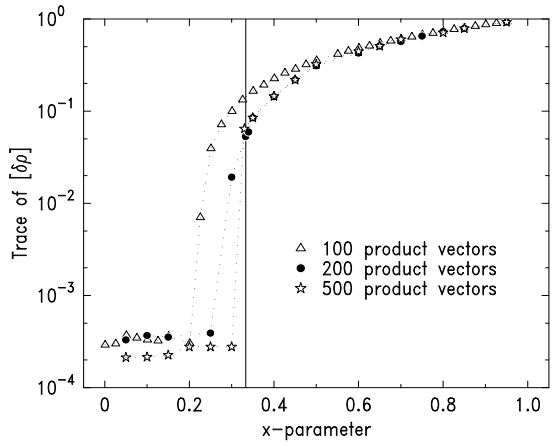}}
\vspace*{0.8cm}
\caption{ The best separable approximation to a Werner state $\rho_{w}$. We plot the the value ${\rm Tr}(\delta\rho)$ for the matrix $\rho_{w}$ characterized by the fraccion of the singlet $x$ (see(\ref{wernma})).
The vertical line indicates the separability border (Eq.(\ref{partialpos})).
(The numerical precision of the algorithm is
set to $10^{-4}$, so that ${\rm Tr}(\delta\rho)$ must be $\geq 10^{-4}$).
\label{dipole}}
\end{figure}
Let us now analyze the "inseparability" properties of $\rho_{w}$.
The matrix $\delta\rho$ when it does not vanish, i.e. for $x>1/3$ corresponds
to the projector onto the maximally entangled singlet  $\ket{\Psi^-}=1/{\sqrt 2}(\ket{\uparrow\downarrow}-\ket{\downarrow\uparrow})$. Thus, a Werner state
of the type $\rho_{w}$  can always be decomposed as:
\begin{equation}
\rho_w(x)=\lambda(x){\rho_s}+(1-\lambda(x))\ket{\Psi^-}\bra{\Psi^-}
\end{equation}
with $\lambda=1$ for $x\leq 1/3$ ( $\Longleftrightarrow \rho_{w}=\ros$), 
and $0\leq\lambda<1$ for $x>1/3$.
A measure of the entanglement of $\rho_w$ is, therefore, naturally provided by the value of the corresponding $\lambda$, i.e. $E(\rho_{w}(x))=(1-\lambda(x))$-ebits,
since the singlet has a value of entanglement of 1 e-bit (see (Eq.(\ref{pentan})). This measure does not coincide 
with other measures of the entanglement of $\rho_w$\cite{bennett2,martin}. A further analysis of this entanglement measure will be presented elsewhere.

Summarizing, we have presented a method to construct the best separable approximation to an arbitrary density matrix of a composite quantum system
(of arbitrary dimensions). The method provides a necessary condition for separability 
of a density matrix. Furthermore, for composite systems of dim[$\Hil$]=4, 
it also provides with unambiguous measure of the entanglement of its non separable states.

It is a pleasure to thank A. Barenco, A. Peres, P. Horodecki, S. Popeuscu and R. Werner for very useful discussions. A.S acknowledges  the Elsag-Baily -I.S.I Foundation meeting on quantum computation 1997 and financial support from European Community.


\begin{references}
\bibitem{notka}  The solution of Eqs. 8 requires to solve inverse equations of the type ($\ket{\phi}=\rho^{-1}{\ket\psi}$) several times. From a numerical point of view, the convergence of these kind of problems is often ill-defined. 
To circumvent this difficulty, we maximize each pair of projection operators $(P_{\alpha},P_{\beta}$) by a trial-error method. Although this slows down the calculation, the program
takes few minuts to calculate BSA for a given set of 100 product vectors.
\end{references}
\end{document}